\documentclass[prl,superscriptaddress,twocolumn,nofootinbib]{revtex4}

\usepackage{amsfonts,amssymb,amsmath}
\usepackage{color}
\usepackage{graphicx}
\usepackage{dcolumn}
\usepackage{bm}
\usepackage{natbib}

\newcommand{\B}[1]{{\mathbb #1}}
\newcommand{\C}[1]{{\mathcal #1}}

\newcommand{\BF}[1]{{\mathbf #1}}

\newcommand{\beq}{\begin{equation}}
\newcommand{\eeq}{\end{equation}}
\newcommand{\bea}{\begin{eqnarray}}
\newcommand{\eea}{\end{eqnarray}}
\newcommand{\nn}{\nonumber}

\newcommand{\Tr}{\mathop{\rm Tr}}

\newcommand{\half}{\frac 12}
\newcommand{\third}{\frac 13}
\newcommand{\quarter}{\frac 14}

\newcommand{\Slash}[1]{{\ooalign{\hfil#1\hfil\crcr\raise.167ex\hbox{/}}}}

\begin{document}


\preprint{arXiv:yymm.nnnn}


\title{Higgs inflation in minimal supersymmetric $SU(5)$ GUT}

\author{Masato Arai}
\email{Masato.Arai(AT)utef.cvut.cz}
\affiliation{Institute of Experimental and Applied Physics,
Czech Technical University in Prague, 
Horsk\' a 3a/22, 128 00 Prague 2, Czech Republic}
\author{Shinsuke Kawai}
\email{kawai(AT)skku.edu}
\affiliation{
Institute for the Early Universe (IEU),
11-1 Daehyun-dong, Seodaemun-gu, Seoul 120-750, Korea} 
\affiliation{Department of Physics, 
Sungkyunkwan University,
Suwon 440-746, Korea}
\author{Nobuchika Okada}
\email{okadan(AT)ua.edu}
\affiliation{
Department of Physics and Astronomy, 
University of Alabama, 
Tuscaloosa, AL35487, USA} 

\date{\today}


\begin{abstract}

The Standard Model Higgs boson with large nonminimal coupling to the gravitational curvature
can drive cosmological inflation.
We study this type of inflationary scenario 
in the context of supersymmetric grand unification and point out that it is naturally implemented in 
the {\em minimal} supersymmetric $SU(5)$ model, and hence virtually in any GUT models.
It is shown that with an appropriate K\"{a}hler potential the inflaton trajectory settles down to the 
Standard Model vacuum at the end of the slow roll.
The predicted cosmological parameters are also consistent with the 7-year WMAP data.
\end{abstract}


\pacs{98.80.Cq, 04.65.+e, 12.10.Dm, 12.60.Jv}
\keywords{Inflationary cosmology, grand unified theory, supergravity}
\maketitle


\section{Introduction}
Recently the idea that the Standard Model (SM) Higgs field may be identified with an inflaton 
field, has attracted much attention 
\cite{
BS,Barvinsky:2008ia,DeSimone:2008ei,BKKSS:2009,Barbon:2009ya,Bezrukov:2010jz,Burgess,Hertzberg:2010dc,LernerMcDonald}.
The major r\^{o}le is played by the nonminimal coupling to gravity, which renders  
the Higgs mass to be within the range of $126 - 194$ GeV 
\cite{BS,Barvinsky:2008ia,DeSimone:2008ei,BKKSS:2009},
while keeping the amplitude of the primordial curvature perturbation at the scale of
$\sim 10^{-5}$.
The idea of inflation by nonminimally coupled inflaton field itself is certainly not new 
\cite{nonminimal}.
Nevertheless, the striking agreement with the present-day cosmological data, combined with the 
minimalistic nature of the model, makes this type of scenario very attractive. 
The predicted mass range of the Higgs particle is also interesting for the physics of
the Large Hadron Collider.

The Higgs potential in the SM is unstable against quantum corrections 
(the hierarchy problem) and it therefore is reasonable to reconsider Higgs inflation in 
supersymmetric theory 
 \cite{Einhorn:2009bh,FKLMvP}.
It is shown in \cite{Einhorn:2009bh} that Higgs inflation cannot be implemented within the 
minimal supersymmetric Standard Model (MSSM), as the field content of the latter is too restrictive.
Instead, with an extra field (i.e. in the next-to-minimal supersymmetric Standard Model, NMSSM) 
a sensible scenario of Higgs inflation is found to be possible.
The NMSSM model has tachyonic instability in the direction of the extra field,
but this can be cured by considering a noncanonical K\"{a}hler potential 
 \cite{FKLMvP}.

In this paper we discuss the possibility of Higgs inflation in supersymmetric grand unified 
theory (GUT).
There are several reasons to motivate this study.
One obvious reason is that the energy scale of inflation is typically above the grand unification
scale, and it is unnatural to suppose that the SM Lagrangian is valid all the 
way up to the scale of inflation;
as the GUT scale destabilises the electroweak scale without supersymmetry, 
it seems that supersymmetric GUT is an appropriate theory to start with.
Another reason is the puzzling necessity of the extra field besides the MSSM fields 
for successful Higgs inflation, as alluded to above;
going beyond the MSSM is somewhat against the minimalistic guiding principle of the original 
Higgs inflation, and as the NMSSM is structurally similar to the $SU(5)$ GUT model, it seems 
natural to conjecture that the $SU(5)$ GUT, rather than the NMSSM, may be a more appropriate 
minimal supersymmetric theory that accommodates Higgs inflation. 
Obvious questions are then whether it is possible to obtain enough inflation (e-folding) somewhere 
between the Planck scale and the GUT scale, and if so whether the prediction of the cosmological
parameters is consistent with the present observation.
We shall address these issues below, and find that a viable Higgs inflationary scenario
nicely fits into the minimal $SU(5)$ model.
We shall employ supergravity embedding of GUT \cite{sGUT},
since the nonminimal coupling of the Higgs field to gravity naturally arises in that framework. 

\begin{figure*}
\begin{eqnarray*}
\begin{array}{ccccc}
\includegraphics[width=60mm]{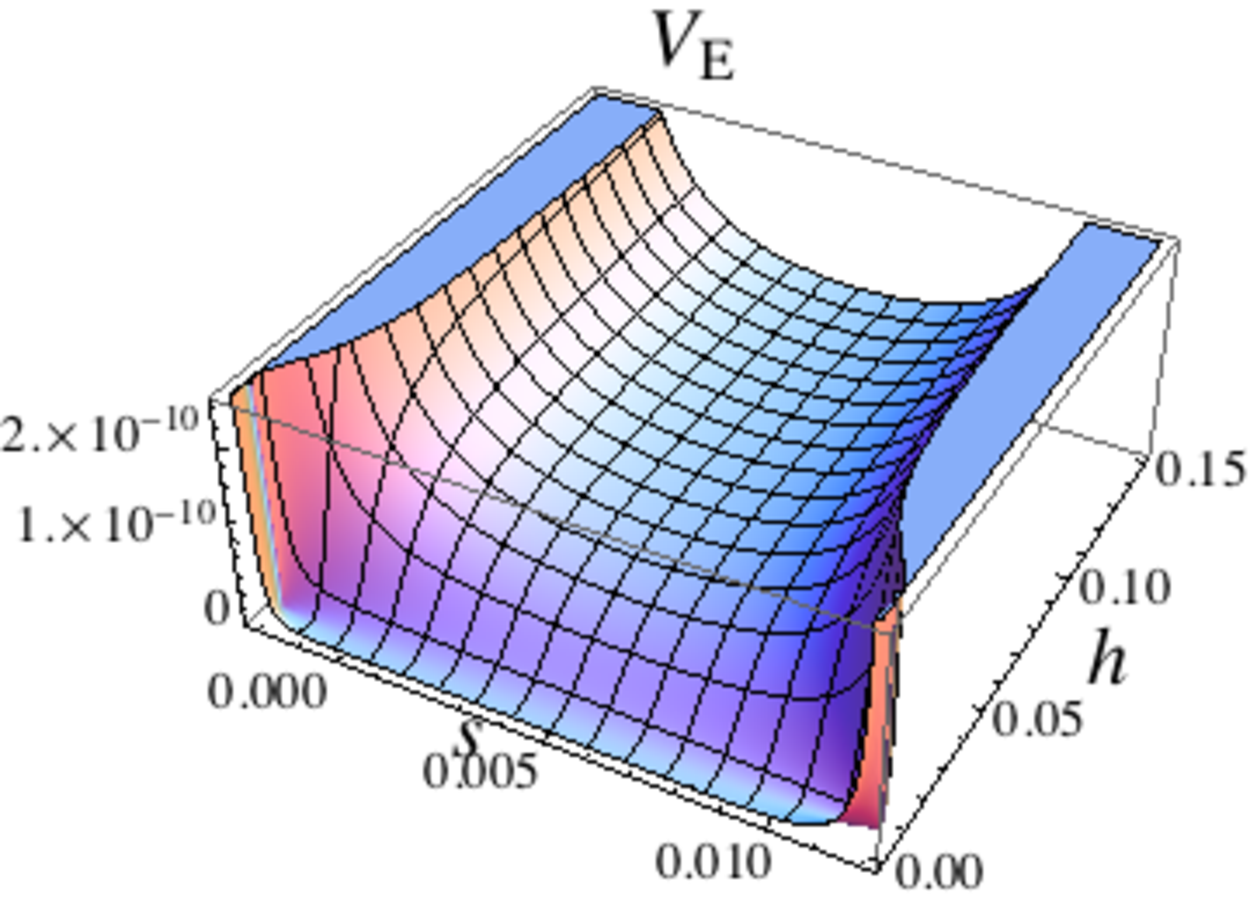}
&~~~&
\includegraphics[width=40mm]{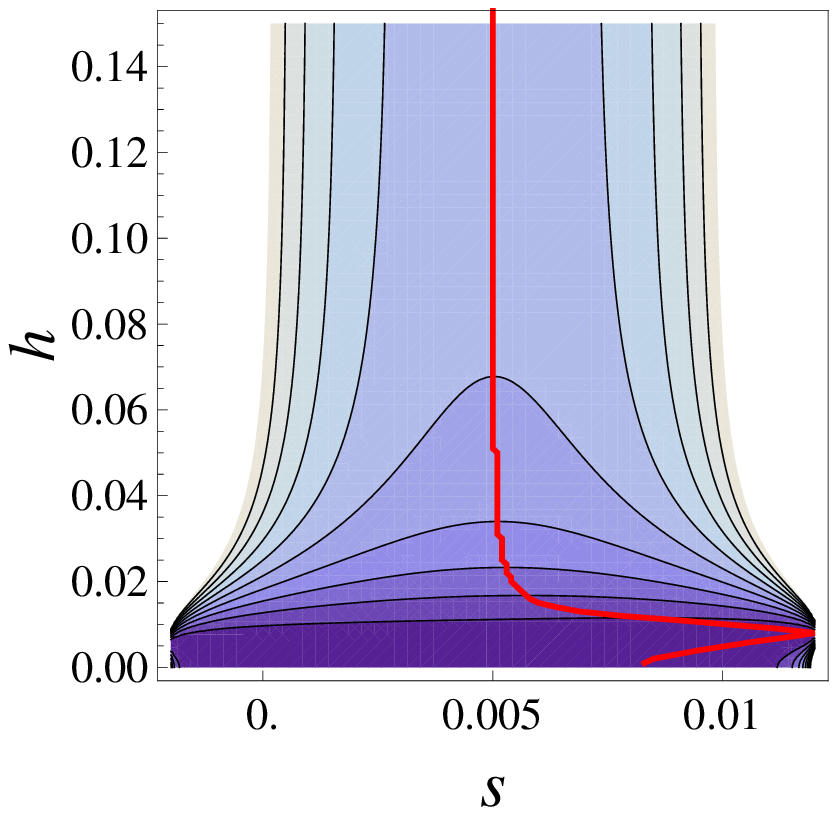}
&~~~&
\includegraphics[width=60mm]{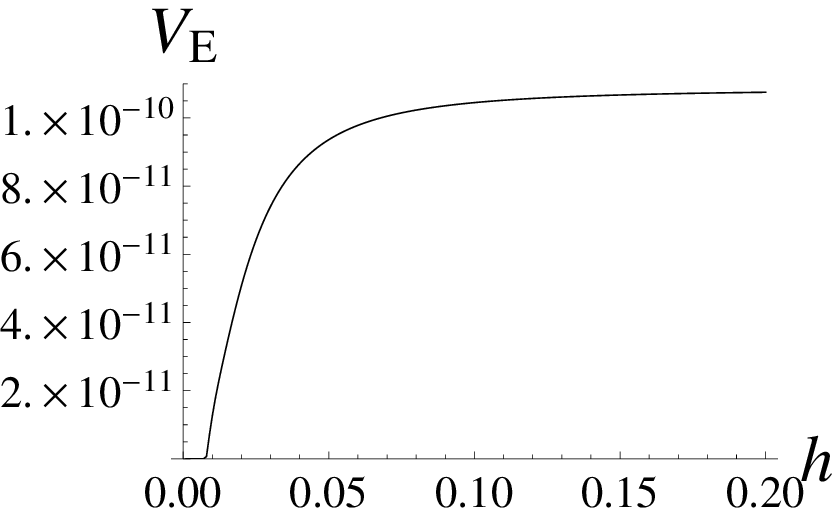}
\end{array}
\end{eqnarray*}
\caption{\label{fig:VE}
The scalar potential $V_{\rm E}$ in the Einstein frame (left), 
the inflaton trajectory in the contour plot of the same potential (middle), 
and the minima of the scalar potential $V(s(h),h)$ plotted against $h$ (right). 
In the middle panel the thick red curve is the inflaton trajectory.
We have chosen $\rho=0.5$, $\lambda=0.5$, $\omega=-100$, $\zeta=10000$.
The nonminimal coupling $\gamma=1.86\times 10^4$ is fixed by the amplitude of the curvature 
perturbation, evaluated for e-folding $N_{\rm e}=60$.
}
\end{figure*}

\section{Supersymmetric $SU(5)$ GUT}
The minimal supersymmetric $SU(5)$ model consists of a vector supermultiplet 
transforming as an adjoint ${\bf 24}$ of the $SU(5)$, as well as 5 types of chiral 
supermultiplets, 
namely $N_{\rm f}$ (the number of flavours) multiplets in ${\bf \bar 5}$ 
(that include $\bar d$ and $L$ of the MSSM), $N_{\rm f}$ multiplets in ${\bf 10}$ 
(include $Q$, $\bar u$, and $\bar e$), one each in ${\bf 24}$ (denoted $\Sigma$), 
${\BF 5}$ ($H$) and ${\bf \bar 5}$ ($\overline{H}$).
$\Sigma$ is the Higgs multiplet responsible for breaking the GUT symmetry, 
while $H$ and $\overline{H}$ respectively include the up- and down-type MSSM Higgs multiplets. 
Among these, only the three Higgs chiral multiplets $\Sigma$, $H$ and $\overline{H}$ 
play r\^{o}les in the dynamics of inflation.
We shall hence disregard the other fields.
The superpotential of our model is,
\beq
W=\overline{H}\left(\mu+\rho\Sigma\right)H
+\frac{m}{2}\Tr(\Sigma^2)+\frac{\lambda}{3}\Tr(\Sigma^3),
\label{eqn:Super}
\eeq
and the K\"{a}hler potential is $K=-3 \Phi$, with
\bea
\Phi&=&1-\third\left(\Tr\Sigma^\dag\Sigma+\vert H\vert^2+\vert\overline{H}\vert^2\right)
-\frac{\gamma}{2}\left(\overline{H}H+H^\dag\overline{H}^\dag\right)\nn\\
&&\qquad
+\frac{\tilde\omega}{3}\left(\Tr\Sigma^\dag\Sigma^2+\Tr\Sigma^{\dag 2}\Sigma\right)
+\frac{\zeta}{3}\left(\Tr\Sigma^\dag\Sigma\right)^2,\label{eqn:Kahler}
\eea
where $\mu$, $\rho$, $m$, $\lambda$, $\gamma$, $\zeta$, $\tilde\omega$ are constant
parameters (for simplicity we assume them to be real).
The cubic and the quartic terms have been included in the K\"{a}hler potential, for reasons to be
discussed shortly.
We shall set the reduced Planck scale $M_{\rm P}=2.44\times 10^{18}$ GeV to unity.

For the model to be phenomenologically consistent, the $SU(5)$ symmetry needs to be broken 
down to the SM gauge group $SU(3)\times SU(2)\times U(1)$.
This is accomplished as usual by setting,
\beq
\Sigma=\sqrt{\frac{2}{15}} S\ {\rm diag}\left(1, 1, 1, -\frac 32, -\frac 32\right),
\label{eqn:Sigma}
\eeq
with $S$ a chiral superfield.
The MSSM Higgs doublets $H_u$, $H_d$ and the Higgs colour triplets
$H_c$, $\overline{H}_c$ are embedded in $H$ and $\overline{H}$ as
\beq
H=\left(\begin{array}{c}
         H_c\\ H_u
        \end{array}\right),
\qquad
\overline{H}=\left(\begin{array}{c}
         \overline{H}_c\\ H_d
        \end{array}\right).
\label{eqn:XY}\eeq
The superpotential now reads
\bea
&&W
=\left(\mu+\sqrt{\frac{2}{15}}\rho S\right)\overline{H}_c H_c
+\left(\mu-\sqrt{\frac{3}{10}}\rho S\right)H_uH_d\nn\\
&&\qquad+\frac m2 S^2-\frac{\lambda}{3\sqrt{30}}S^3.
\label{eqn:W1}
\eea
The masses of $H_u$ and $H_d$ are in the electroweak scale, which is negligibly smaller
than the typical scale $M_{\rm P}$ of the inflationary dynamics. 
Thus the expectation value of the second term in (\ref{eqn:W1}) must vanish,
$\mu=\sqrt{3/10}~\rho \langle S\rangle$, 
where 
$\langle S\rangle= v\equiv 2\times 10^{16} \mbox{~GeV}$
is the GUT scale.
The first term of (\ref{eqn:W1}) indicates that $H_c$ and $\overline{H}_c$ have GUT scale masses.
For the colour symmetry to be unbroken we require that they are already stabilised at
$\langle H_c\rangle =\langle\overline{H}_c\rangle=0$,
from the onset of the inflation.
During inflation the dominant r\^{o}le is played by the MSSM Higgs fields $H_u$ and $H_d$, 
which settle down to the present values after the inflation.
When $H_u$, $H_d\ll 1$ 
(i.e. close to the end of inflation) the stationary condition
$\delta W/\delta S=0$
with $H_c= \overline{H}_c= 0$ yields
$S(m-\lambda S/\sqrt{30})=0$.
Since the GUT symmetry must be broken, $\langle S\rangle=v\neq 0$ and we must have
$m
=\frac{\lambda}{\sqrt{30}} v$.
The charged Higgs can be consistently set to be zero,
\beq
H_u=\left(\begin{array}{c} 0\\ H_u^0\end{array}\right),
\quad
H_d=\left(\begin{array}{c} H_d^0\\ 0\end{array}\right),
\eeq
and parametrizing
$S=s e^{i\alpha}$,
$H_u^0=\frac{1}{\sqrt 2}h_1 e^{i\alpha_1}$,
$H_d^0=\frac{1}{\sqrt 2}h_2 e^{i\alpha_2}$,
with $s,h_1,h_2,\alpha,\alpha_1,\alpha_2\in{\B R}$, and further setting
$h_1=h\sin\beta$ and $h_2=h\cos\beta$, 
the model depends on five parameters $\rho$, $\lambda$, $\gamma$, $\tilde\omega$, $\zeta$, 
and six real scalar fields $s$, $h$, $\alpha$, $\beta$, $\alpha_1$, $\alpha_2$.
Note that $\rho$ and $\lambda$ are parameters appearing in the GUT superpotential and
are typically of order one, while there is no such restriction for
$\gamma$, $\tilde\omega$, and $\zeta$.
Analysing the scalar potential, we find stability at $\alpha=\alpha_1=\alpha_2=0$.
Furthermore, the D-flat condition sets the value of $\beta$ to be $\pi/4$.
Thus the model reduces to a system of two real scalars $h$ and $s$, with the scalar-gravity part of 
the Jordan frame Lagrangian (cf. 
\cite{FKLMvP}),
\beq
{\C L}_{\rm J}=
\sqrt{-g_{\rm J}}\Big[
\half\Phi R_{\rm J}-\half g_{\rm J}^{\mu\nu}\partial_\mu h\partial_\nu h
-\kappa g_{\rm J}^{\mu\nu}\partial_\mu s\partial_\nu s
-V_{\rm J}\Big].
\label{eqn:LJ}
\eeq
The subscript J denotes quantities in the Jordan frame,
$\kappa\equiv K_{SS^\dag}=1-4\omega s-4\zeta s^2$ is the nontrivial component of the 
K\"{a}hler metric,
$\omega\equiv-\tilde\omega/\sqrt{30}$, and 
\beq
\Phi=1-\frac{1}{3}s^2+\frac{2 \omega}{3} s^3+\frac{\zeta}{3}s^4+\left(\frac{\gamma}{4}
-\frac 16\right) h^2.
\eeq
$V_{\rm J}$ is the F-term scalar potential in  the Jordan frame, computed in the standard way
\cite{superconformal}, as
\bea
V_{\rm J}
&=&
\frac{3}{10}\left\{\frac{\rho^2}{2}(s-v)^2h^2
+\frac{1}{\kappa}\left[\frac{\rho}{4}h^2-\frac{\lambda}{3} s(s-v)\right]^2\right\}\nn\\
&&\hspace{-18mm}-\frac{\left\{\frac{2\zeta s+\omega}{\kappa}\!\left[\frac{\rho h^2}{4}
\! -\!\frac{\lambda s(s-v)}{3}\right]\! s^2\!
+\frac{\rho v h^2}{4}\! -\!\frac{\lambda v s^2}{6}
\! -\!\frac{3\gamma\rho h^2 (s-v)}{4}\right\}^2}
{10\left[1+\frac{\gamma}{4}(\frac 32 \gamma-1) h^2+\frac{\zeta+\omega^2}{3\kappa} s^4\right]}.
\eea

\section{The inflation dynamics}\label{sec:InfDyn}
The dynamics of inflation is encoded in the scalar potential
$V_{\rm E}=\Phi^{-2}V_{\rm J}$ in the Einstein frame.
If we take the canonical form of the K\"{a}hler potential (i.e. $\omega=\zeta=0$), the potential
exhibits tachyonic instability in the direction of the $s$-field. 
Just as in the case of the NMSSM Higgs inflation 
 \cite{Einhorn:2009bh,FKLMvP}
the instability is controlled by 
introducing a quartic term ($\zeta\neq 0$) in the K\"{a}hler potential. 
In the GUT model, however, this is not the whole story, as the quartic term has a serious side effect:
the SM vacuum becomes disfavoured and the $SU(5)$ symmetry tends to be 
restored at the end of inflation.
This problem is resolved by allowing a cubic term\footnote{Higher (say sextic) terms in the 
K\"{a}hler potential can also solve this problem. }, $\omega\neq 0$. 
Note that these terms are perfectly consistent with the supergravity embedding.
The bottom line is that for a wide range of the parameter space with up to quartic order terms 
in the K\"{a}hler potential, there exist reasonable trajectories of the inflaton field.
In Fig.\ref{fig:VE} we show the shape of the scalar potential $V_{\rm E}$ (the left panel), 
the inflaton trajectory (centre), and the values of $V_{\rm E}$ at local minima (bottom of the valley)
for given $h$ (right).
In this example we have taken $\rho=\lambda=0.5$, $\omega=-100$, $\zeta=10000$, and 
$\gamma=1.86\times 10^4$ 
(this value of $\gamma$ is determined for the e-folding number $N_{\rm e}=60$, 
as discussed below).
The plateau of the potential at the large $h$ values is a characteristic feature of Higgs inflation.
As the field $s$ controls breaking of the GUT symmetry, the trajectory shows that $SU(5)$ is 
broken from the onset, indicating that problematic topological defects are not produced during
inflation.
For this parameter set the dynamics of the slow roll inflation is dominated by
the $h$ field, as the displacement of $s$ is negligibly small ($\Delta\tilde s/\Delta h \lesssim$ 2\%,
with suitable normalisation $d\tilde s=\sqrt{2\kappa}ds$).
Assuming that $s$ is nearly constant\footnote{
The value of $s=s(h)$ is taken at the local minimum of $V_{\rm E}$ for a given $h$,
and derivatives of $s$ are set to be zero.
}, the model simplifies to single field inflation.
The Lagrangian (\ref{eqn:LJ}) can then be written in a form similar to the SM Higgs inflation
\cite{
BS,Barvinsky:2008ia,DeSimone:2008ei,BKKSS:2009,Barbon:2009ya,Burgess,Hertzberg:2010dc,LernerMcDonald},
\beq
{\C L}_{\rm J}=\sqrt{-g_{\rm J}}\left[\frac{M^2+\xi h^2}{2} R_{\rm J}
-\half g_{\rm J}^{\mu\nu}\partial_\mu h\partial_\nu h-V_{\rm J}\right],
\eeq
with $M^2=1-\third s^2+\frac{2}{3}\omega s^3+\third\zeta s^4$ and $\xi=\quarter\gamma-\frac 16$.

\section{Cosmological parameters}\label{sec:Cos}
The slow roll parameters,
\beq
\epsilon=\half\left(\frac{1}{V_{\rm E}}\frac{d V_{\rm E}}{d\hat h}\right)^2,
\qquad 
\eta=\frac{1}{V_{\rm E}}\frac{d^2 V_{\rm E}}{d\hat h^2},
\eeq
are defined for the scalar potential $V_{\rm E}$ and the canonically normalised inflaton field 
$\hat h$ in the Einstein frame. 
The latter is related to $h$ by
\beq
d\hat h=\frac{\sqrt{M^2+\xi h^2+6\xi^2 h^2}}{M^2+\xi h^2} dh.
\eeq
For given $(\lambda, \rho, \omega, \zeta)$, the nonminimal coupling $\xi$ is determined from the 
power spectrum of the curvature perturbation ${\C P}_R=V_{\rm E}/24\pi^2\epsilon$.
The slow roll terminates when either of the slow roll parameters ($\epsilon$ in the present case) 
becomes ${\C O}(1)$.
The values of the inflaton $h=h_*$ at the end of the slow roll and $h_k$ at 
the horizon exit of the comoving CMB scale $k$, are related by the e-folding number 
$N_{\rm e}=\int_{h_*}^{h_k}dh V_{\rm E}({d\hat h}/{dh})/({d V_{\rm E}}/{d\hat h})$.
At $h=h_k$ the shape of $V_{\rm E}$ is constrained by the power spectrum ${\C P}_R$.
We have used the maximum likelihood value $\Delta^2_R(k_0)=2.42\times 10^{-9}$
from the 7-year WMAP data \cite{Komatsu:2010fb}, where 
$\Delta_R^2(k)=\frac{k^3}{2\pi^2}{\C P}_R(k)$ and the normalisation is fixed at 
$k_0=0.002 \mbox{ Mpc}^{-1}$.
With $\lambda=\rho=0.5$, $\omega=-100$ and $\zeta=10000$, we find $h_*=0.0146$, 
$h_k=0.128$ and $\xi=4646$ for $N_{\rm e}=60$.
For $N_{\rm e}=50$ we obtain $h_*=0.0160$, $h_k=0.130$ and $\xi=3895$.
With these parameters the prediction of the scalar 
spectral index $n_s\equiv d\ln{\C P}_R/d\ln k=1-6\epsilon+2\eta$ and the tensor-to-scalar ratio
$r\equiv {\C P}_{\rm gw}/{\C P}_{\rm R}=16\epsilon$ can be evaluated.
We find $n_s=0.968$, $r=0.00296$ for $N_{\rm e}=60$, and 
$n_s=0.962$, $r=0.00419$ for $N_{\rm e}=50$.
These results are shown in Fig.{\ref{fig:rns_WMAP7_sGUT} with observational constraints 
\cite{Komatsu:2010fb}.
The prediction for $n_s$ and $r$ is insensitive to the change of 
$\lambda$ and $\rho$, as long as they are ${\C O}(1)$.
With $(N_{\rm e}, \lambda, \rho)=(60, 0.1, 0.5)$ and $(60, 0.5, 0.1)$, for example, 
we obtain the same prediction $n_s=0.968$ and $r=0.00296$ as above .
In contrast to the nonsupersymmetric case, the inflationary dynamics does not constrain the 
Higgs mass at the electroweak scale.

\begin{figure}
\includegraphics[width=80mm]{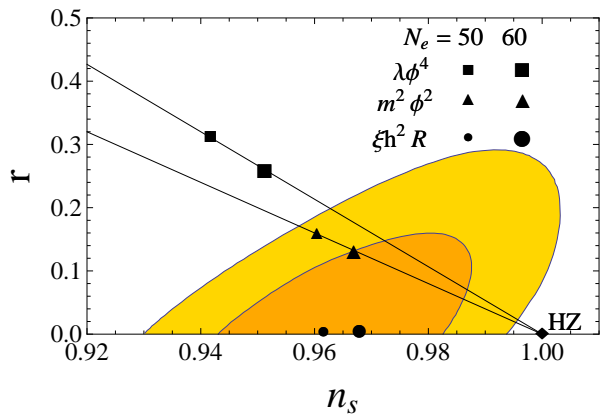}
\caption{The tensor-to-scalar ratio $r$ and the scalar spectral index $n_s$, with 
the 68\% and 95\% confidence level contours from the WMAP7$+$BAO$+H_0$ data
\cite{Komatsu:2010fb}.
The Harrison-Zel'dovich (HZ) values as well as the predictions of the $\phi^4$ and $\phi^2$ 
chaotic inflation models are also shown for comparison.
\label{fig:rns_WMAP7_sGUT}
}
\end{figure}

\section{Discussion}\label{sec:Discussion}
In this paper we have discussed Higgs inflation in supersymmetric GUT, taking the minimal 
$SU(5)$ model as a concrete example. 
In the early days the proposals of cosmological inflation were made for the Higgs field in the 
GUT models \cite{inflation}. 
It is intriguing to see that the prediction based on the simplest GUT, 
with the help of nonminimal coupling to gravity, 
is in perfect fit with today's observational constraints. 

The nonminimal coupling is consistent with the symmetries of general relativity and
the SM, and it naturally arises in quantum field theory in curved spacetime 
\cite{Birrell:1982ix}. 
The value of the coupling $\xi\sim 10^4$, however, is rather large.
This is a generic feature of Higgs inflation, since successful slow-roll requires 
$h^2\lesssim M_{\rm P}^2\lesssim\xi h^2$ \cite{BS}.
It has been argued that such large nonminimal coupling could violate the unitarity bound,
since the cut-off scale evaluated as $M_{\rm P}/\xi$ is considerably lower than the Planck scale
\cite{Barbon:2009ya,Burgess,LernerMcDonald,Hertzberg:2010dc}.
Others contend that such a criticism is not valid, arguing that at large field values
$\gtrsim M_{\rm P}/\xi$ the cut-off scale is actually field-dependent 
\cite{BKKSS:2009,Bezrukov:2010jz,FKLMvP}.
The large nonminimal coupling is, at any rate, a key feature of the Higgs inflation and it is certainly
worthwhile understanding possible dangers arising from this.
%
%
Another type of criticism concerns the quantum stability of the classical potential.
This problem was studied using renormalisation group (RG) analysis 
\cite{BKKSS:2009,Barvinsky:2008ia,DeSimone:2008ei}, and the effects of renormalisation 
are found to be small except for some extreme values of parameters. 
We have also performed RG analysis in our GUT model and verified that the effects are
small (less than $3\%$ for $r$, less than $2\%$ for $\xi$, and less than $0.1\%$ for $n_s$).
This is expected, since inflation takes place in a narrower energy range of $10^{16} - 10^{18}$ 
GeV and the RG effects should be smaller than the SM case.

A closer look at the potential $V_{\rm E}$ shows that its minimum is at a small negative value,
$\sim -2 \times 10^{-16} M_{\rm P}^4$, for our parameter choices.  
This is offset by a contribution from the supersymmetry breaking sector
and the scenario does not suffer from the cosmological constant problem.
In our scenario the energy scale of inflation is in the GUT scale and the Higgs fields are directly 
coupled to the SM particles.
This indicates that the reheating temperature is high, typically from the intermediate 
to the GUT scale.
It would be interesting to discuss further phenomenological implications, 
such as the gravitino problem and baryogenesis.

In this paper we considered a single-field Higgs inflation model appropriate for our parameter 
choice $\zeta=10000$, $\omega=-100$ of the K\"{a}hler potential.
These values are not too exotic, as $\langle\Phi\rangle$ is still very close to $1$ and the Planck 
scale after inflation is nearly $M_{\rm P}$.
For smaller values of $\zeta$ and $|\omega|$, the displacement of $s$ during inflation becomes 
large.
This leads to two-field inflation, which is also of interest, in particular, due to 
possible generation of detectable large non-Gaussianity. 
Supersymmetric models of Higgs inflation necessarily involve multiple fields 
\cite{Einhorn:2009bh}.
The engendered isocurvature mode can, in principle, distinguish various models of Higgs inflation.

Finally, the scenario can also be extended to other GUT models whose gauge group contains 
$SU(5)$ as a subgroup.
When the Higgs multiplets of the GUT model contain ${\bf 5}$, ${\bf \bar{5}}$ and ${\bf 24}$
of the minimal $SU(5)$ GUT, a superpotential like (\ref{eqn:Super}) can be introduced.
Then a viable model of Higgs inflation is implemented, as described in this Letter. 
One such simple example is the $SO(10)$ GUT with Higgs multiplets
in ${\bf 10}$ and ${\bf 54}$ representations.

\subsection*{Acknowledgments}
This work was supported in part by the Research Program MSM6840770029, 
ATLAS-CERN International Cooperation (M.A.), the WCU Grant No. R32-2008-000-10130-0 (S.K.), 
and by the DOE Grant No. DE-FG02-10ER41714 (N.O.).

\bigskip




\end{document}